\documentclass[12pt, a4paper]{article}
\pdfoutput=1

\usepackage{amsmath}
\usepackage{amsfonts}
\usepackage{amssymb}
\usepackage{graphicx, rotating}
\usepackage{epsfig}
\usepackage{latexsym}
\usepackage{graphicx}
\usepackage{color}
\usepackage{amsmath,bm,amssymb}
\usepackage{cite}
\usepackage{slashed}
\usepackage{hyperref}
\hypersetup{colorlinks, citecolor=bluscuro, linkcolor=black, urlcolor=bluscuro}
\definecolor{rossos}{cmyk}{0,1,1,0.55}
\definecolor{bluscuro}{rgb}{0.15, 0.2, .85}
\definecolor{bluchiaro}{cmyk}{1,.3,0.,0.1}

\newcommand{\be}{\begin{equation}}
\newcommand{\ee}{\end{equation}}
\newcommand{\bea}{\begin{eqnarray}}
\newcommand{\eea}{\end{eqnarray}}

\newcommand{\arXiv}[2]{\href{http://arxiv.org/pdf/#1}{{\tt [#2/#1]}}}

\def\bma#1{\mbox{\boldmath{$#1$}}}

\begin{document}
\allowdisplaybreaks
\begin{titlepage}
\begin{flushright}
CERN-TH-2023-075
\end{flushright}
\vspace{.3in}

\vspace{1cm}
\begin{center}
{\Large\bf\color{black} 
Thick Wall from Thin Walls} \\
\vspace{1cm}{
{\large J.R.~Espinosa}
\vspace{0.3cm}
} \\[7mm]
{\it {Instituto de F\'{\i}sica Te\'orica, IFT-UAM/CSIC, \\ 
C/ Nicol\'as Cabrera 13-15, Campus de Cantoblanco, 28049, Madrid, Spain}}\\
{\it and}\\
{\it {Theoretical Physics Department, CERN, CH-1211 Geneva 23, Switzerland}}
\end{center}
\bigskip

\vspace{.4cm}

\begin{abstract}
An Euclidean bounce describing vacuum decay can be considered as an infinite stack of concentric thin shells to which a thin-wall action can be assigned. The integral over all shells produces then a tunneling action  that is precisely the action functional in field space  of the so-called tunneling potential formalism. This procedure, which works also when gravity is included, gives the simplest derivation of such actions.

\end{abstract}
\bigskip

\end{titlepage}

\section{Introduction \label{sec:intro}} 

In the problem of false vacuum decay in quantum field theory, there is one particular situation that can be treated semi-analytically and simply: when the energy difference between the false and true vacua is small compared to other scales. In this case, the scalar field $\phi(r)$ of the  Euclidean bounce that describes the decay makes a sharp transition between the vacua at a well-defined bounce radius (thus the name ``thin-wall case'' \cite{Coleman}). However, in general, the bounce has a thick wall, and no analytical treatment is possible.

A generic bounce can be thought of as an infinite stack of concentric thin-wall shells of infinitesimal width (from $\phi$
to $\phi+d\phi$) across which the (Euclidean) energy changes infinitesimally. As shown in this paper, one can associate an infinitesimal action to such
slices, given by the simple thin-wall expression, generalized in the appropriate way. Integrating over all slices one recovers the tunneling action for the general case, albeit written in a particular form: the one recently proposed  in the tunneling potential formalism \cite{E,Eg}.   The next section shows how this is done (also with gravity) and the appendix extends the derivation to general spacetime dimension $d>2$. 

The tunneling potential formalism \cite{E,Eg} for the calculation of tunneling actions does not rely on Euclidean bounces and reformulates the calculation as a simple variational problem in field space. Instead of a bounce one finds a ``tunneling potential'' function, $V_t(\phi)$, that connects the false vacuum and (the basin of) the true vacuum and minimizes an action functional $S[V_t]$, an integral in field space of a simple action density. The resulting action reproduces the Euclidean result and the formalism has a number of appealing properties that have been discussed elsewhere. 

The action functional $S[V_t]$ was first derived starting from the Euclidean approach, using the relation $V_t=V-\dot\phi^2/2$ to get rid of all Euclidean quantities in favor of $V_t(\phi)$ and its derivatives, and arriving at a particular second-order differential equation that $V_t$ should satisfy. Then $S[V_t]$ is  obtained so that its minimization leads to that differential equation for $V_t$ \cite{E,Eg} and its normalization is right to reproduce the Euclidean action. An alternative (simpler) derivation of the $V_t$ action uses a canonical transformation between Euclidean and $V_t$ formalisms \cite{EJK}. The derivation of $S[V_t]$ presented now in this paper is the simplest.

\section{Thick Wall as Infinite Stack of Thin Walls \label{sec:onion}} 
\subsection{No Gravity}
The thin-wall tunneling action ($d=4$ spacetime dimensions, no gravity) is
\be
S_{\rm thin}=\frac{27\pi^2\sigma^4}{2\epsilon^3}\ ,
\ee
where $\sigma$ is the wall tension and $\epsilon>0$ the difference in energy between the two vacua. In the Euclidean formalism \cite{Coleman} one has [false vacuum at $\phi_+$, true vacuum at $\phi_-$, with $V_\pm\equiv V(\phi_\pm)$]
\be
\sigma\simeq\int_{\phi_+}^{\phi_-}\sqrt{2(V-V_-)}d\phi\ ,\quad\quad
\epsilon\simeq -\Delta V\equiv -V_-+V_+>0\ .
\label{thinparamE}
\ee

\begin{figure}[t!]
\begin{center}
\includegraphics[width=0.46\textwidth]{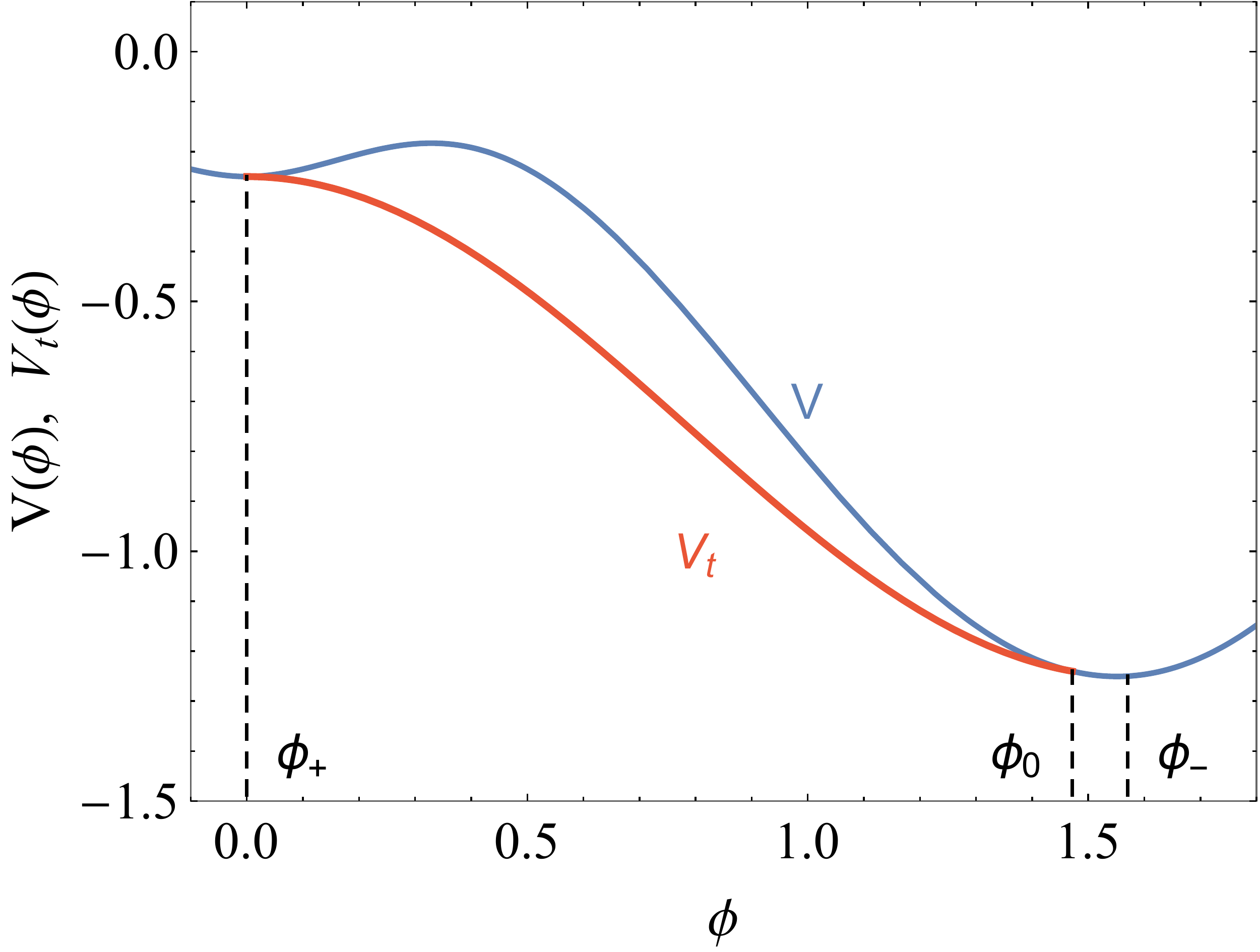}\,\,
\includegraphics[width=0.45\textwidth]{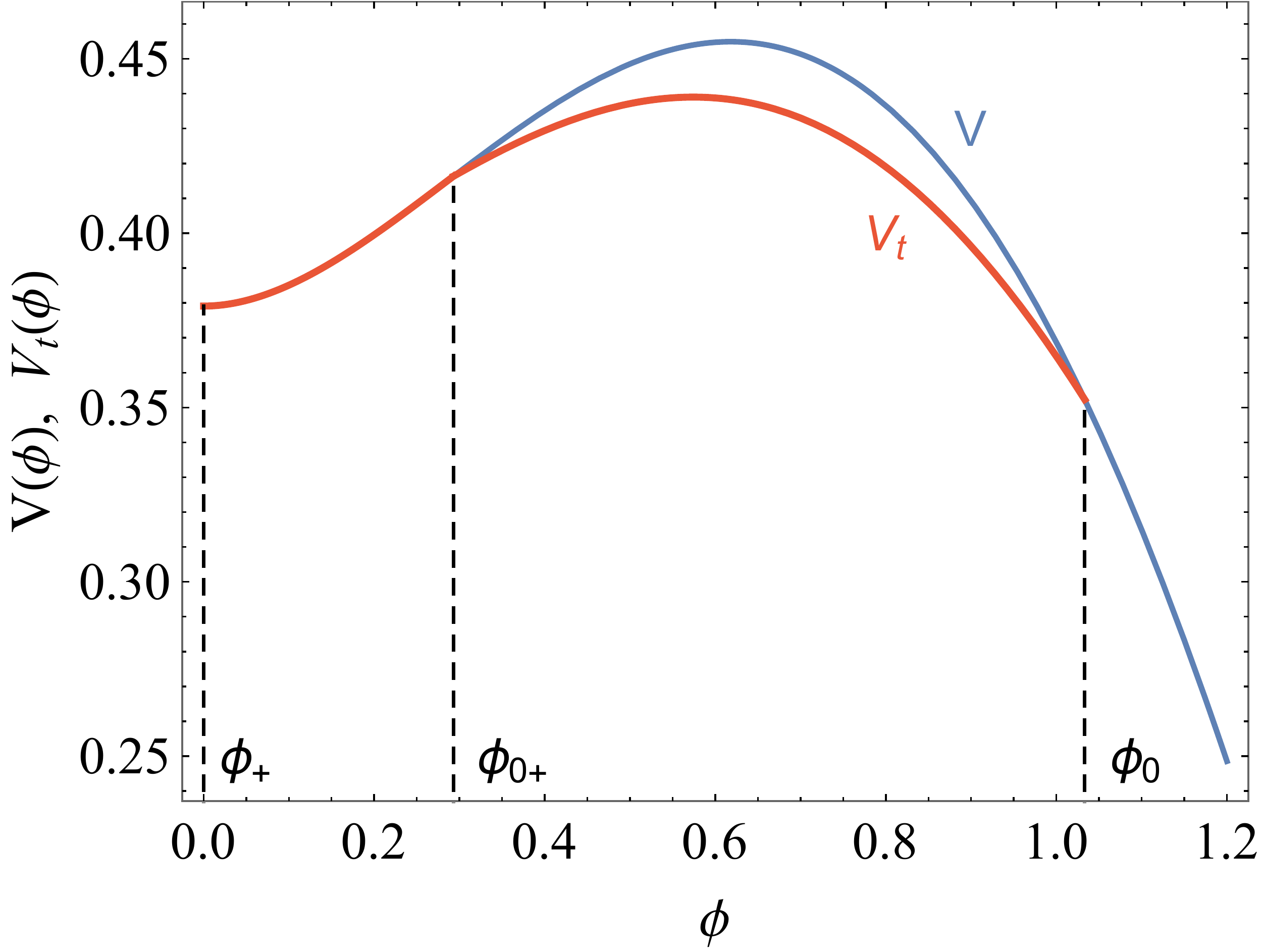}
\end{center}
\caption{Examples of tunneling potentials $V_t$ for AdS decay (left) and dS decay (right). For Minkowski or no-gravity cases, $V_t$ is monotonic and  qualitatively similar to the AdS case.
\label{fig:VtV}
}
\end{figure}

 In the $V_t$ formalism, $\sigma$ and $\epsilon$ are defined more precisely, as
\be
\sigma=\int_{\phi_+}^{\phi_0}\sqrt{2(V-V_t)}d\phi\ ,\quad\quad
\epsilon=-\Delta V_t\equiv -V_{t}(\phi_0)+V_t(\phi_+)>0\ ,
\label{thinparam}
\ee
where $\phi_0$ is the ``exit'' point for tunneling [$\phi_0=\phi(0)$ in the Euclidean formalism] with $\phi_0\simeq\phi_-$ in the thin-wall case. In (\ref{thinparam}), $\epsilon$ is written in terms of $V_t$ rather than $V$, using the fact that $V_t=V$ at $\phi_+$ and $\phi_0$ (points at which $\dot\phi=0$). In $\sigma$,  $V_t(\phi)$ is  the solution of the corresponding differential equation. The left plot in Fig.~\ref{fig:VtV} shows an example of $V_t$. In the thin-wall case, however, $V_t$ can be approximated well by a nearly flat monotonic function that connects the two vacua \cite{E}. 

Now consider any (generally thick-wall) bounce as a collection of concentric thin slices  from $\phi_+$ to $\phi_0$. Through each of them
the field and $V_t$ are changing a little and we assign to them a differential wall tension and differential change 
in (Euclidean) energy as
\be
\boxed{
d\sigma = \sqrt{2(V-V_t)}d\phi\ ,\quad \quad
d\epsilon = -V_t'd\phi\ .}
\label{dthin}
\ee 
The relation $V_t=V-\dot\phi^2/2$ tells us that $V_t$ can also be interpreted as (minus) the Euclidean energy, so that $d\epsilon$ is the infinitesimal change in Euclidean energy across the wall.
To get the thick wall action we  integrate the thin-wall action for all slices:
\be
S=\int_{\phi_+}^{\phi_0}dS_{\rm thin}=\int_{\phi_+}^{\phi_0}\frac{27\pi^2d\sigma^4}{2 d\epsilon^3}=54\pi^2\int_{\phi_+}^{\phi_0}\frac{(V-V_t)^2}{(-V_t')^3}d\phi\ ,
\ee
which reproduces exactly the tunneling action in the $V_t$ formalism \cite{E}.

\subsection{Gravity On: AdS/Minkowski Vacua}
If gravity is included, the thin-wall action for the decay of a Minkowski or AdS false vacuum can be written as \cite{CdL,Parke}
\be
S_{\rm thin}=\frac{12\pi^2}{\kappa^2 V}\left.\left(\sqrt{1-3\kappa V/C^2}-1\right)\right|_{\phi_+}^{\phi_0}=\frac{12\pi^2}{\kappa^2 V_t}\left.\left(\sqrt{1-3\kappa V_t/C^2}-1\right)\right|_{\phi_+}^{\phi_0}\ ,
\ee
where $\kappa\equiv1/m_P^2$ and $C^2$ is
\be
C^2=3\kappa V_{t+} + \frac{1}{\sigma^2}\left(\epsilon-\frac34 \kappa \sigma^2\right)^2=
3\kappa V_- +\frac{1}{\sigma^2}\left(\epsilon+\frac34 \kappa\sigma^2\right)^2
\ .
\label{C2}
\ee
In the literature, this formula is written in terms of $V$ rather than
$V_t$ but we use the fact that $V(\phi_+)=V_t(\phi_+)\equiv V_{t+}$ and $V(\phi_0)=V_t(\phi_0)$ to rewrite it in terms of $V_t$, which is the natural choice to make the thick-from-thin connection.
Similarly, $\sigma$ and $\epsilon$ are defined as in (\ref{thinparam}), although now $V_t$ solves an equation of motion that includes gravitational effects \cite{Eg}.

We slice as before a thick wall in an infinite collection of concentric thin walls with $d\sigma$ and $d\epsilon$ given by the same formulas (\ref{dthin}). For a given thin-wall slice with field between $\phi$ and $\phi+d\phi$, $C^2$ takes now the value 
\be
C^2(\phi)=3\kappa V_t +\frac{d\epsilon^2}{d\sigma^2} =\frac{D^2}{2(V-V_t)}\ ,
\label{C}
\ee
where 
\be
D^2\equiv V_t'{}^2+6\kappa (V-V_t)V_t\ .
\ee
From this it follows $\sqrt{1-3\kappa V_t/C^2(\phi)} = -V_t'/D\ .$
For such slice we obtain the infinitesimal action
\be
dS_{\rm thin} =\frac{12\pi^2}{\kappa^2 V_t}\left.\left[\sqrt{1-\frac{3\kappa V_t}{C^2(\phi)}}-1\right]\right|_{V_t(\phi)}^{V_t(\phi+d\phi)}=\frac{6\pi^2}{\kappa^2V_t^2}\left[2-\sqrt{1-\frac{3\kappa V_t}{C^2(\phi)}}-\frac{1}{\sqrt{1-\frac{3\kappa V_t}{C^2(\phi)}}}\right]V_t'd\phi\ .
\ee
Integrating over all slices and using the previous relations, the thick-wall action is
\be
S=\int_{\phi_+}^{\phi_0}dS_{\rm thin}=\frac{6\pi^2}{\kappa^2}\int_{\phi_+}^{\phi_0}\frac{(D+V_t')^2}{DV_t^2}d\phi\ ,
\ee
which reproduces the tunneling action of the $V_t$ formalism \cite{Eg}. Note that $dS_{\rm thin}>0$.

\subsection{Gravity On: dS Vacua}
For the dS case, as is well known \cite{BW,Eg,Parke}, the vacuum decay can be considered as composed of two parts, a Hawking-Moss type transition from the false vacuum field value $\phi_+$ to a field value $\phi_{0+}$ with higher energy and then a Coleman-De Luccia (CdL) type of transition from $\phi_{0+}$ to $\phi_0$. In the $V_t$ formalism, the first part has $V_t=V$ as depicted in Fig.~\ref{fig:VtV}, right plot.  

For the Hawking-Moss part, we first have to derive the ``thin-wall'' approximation for that case. The Hawking-Moss rate is \cite{HM}
\be
S_{\rm HM} = \frac{24\pi^2}{\kappa^2}\left(\frac{1}{V_+}-\frac{1}{V_{\rm top}}\right)\ ,
\ee
where $V_{\rm top}$ is the value of the potential at the top of the barrier that makes the vacuum classically stable. We can talk of a thin-wall HM transition (thin in field space) from $\phi$ to $\phi+d\phi$ simply setting $V_+=V(\phi)$ and $V_{\rm top}=V(\phi+d\phi)$ to get
\be
dS_{\rm HM} =\frac{24\pi^2 V'}{\kappa^2V^2}d\phi= \frac{24\pi^2 V_t'}{\kappa^2V_t^2}d\phi\ ,
\ee
where, for the last equality, we have used that $V_t=V$ in the HM region of the decay. 

For the CdL part of the transition the thin-wall result  \cite{BW,Parke} can be written as
\be
S_{\rm thin}=\frac{12\pi^2}{\kappa^2}\left\{\frac{1}{V_{t+}}\left[1-\frac{1}{C\sigma}\left(\epsilon-\frac{3\kappa \sigma^2}{4}\right)\right]-
\frac{1}{V_{t0}}\left[1-\frac{1}{C\sigma}\left(\epsilon+\frac{3\kappa \sigma^2}{4}\right)\right]\right\}\ ,
\ee
with $C^2$ as in (\ref{C2}).
For the slice between $\phi$ and $\phi+d\phi$
we get $C^2(\phi)$ as in (\ref{C})
for the AdS/Minkowski case. Setting $V_{t+}=V_t(\phi)$, $V_{t0}=V_t(\phi+d\phi)$, and using $d\sigma$ and $d\epsilon$ as in (\ref{dthin}) 
\be
dS_{\rm thin}=\frac{6\pi^2}{\kappa^2}\frac{(D+V_t')^2}{DV_t^2}d\phi\ ,
\ee
which is, again, the correct action density \cite{Eg}.
Adding up all infinitesimal actions from HM and CdL parts we finally get
\be
S=\frac{24\pi^2}{\kappa^2}\int_{\phi_+}^{\phi_{0+}}\frac{V_t'}{V_t^2}d\phi+\frac{6\pi^2}{\kappa^2}\int_{\phi_{0+}}^{\phi_0}\frac{(D+V_t')^2}{DV_t^2}=\frac{6\pi^2}{\kappa^2}\int_{\phi_{+}}^{\phi_0}\frac{(D+V_t')^2}{DV_t^2}\ ,
\ee
in agreement with the action for dS decay in the $V_t$ formalism \cite{Eg}.

\section{Conclusions}
Although the thin wall action is a particular limit of the general tunneling action, this paper shows how the general (thick) action can be recovered from the particular (thin) case: consider any tunneling bounce as an infinite collection of concentric thin shells, apply to them the thin-wall action (defined in a concrete way) and integrate over the full collection.

Obviously, the thin-to-thick connection discussed in this paper does not lead to an algorithm useful to calculate the action or $V_t$ (e.g. numerically) as the thin-wall infinitesimal actions involve the global  $V_t$.

The result is nevertheless interesting because the general action obtained is naturally a functional in field space and corresponds precisely to the action used in the tunneling potential formalism. The derivation in this paper can thus be considered as the simplest derivation of such actions.

Alternatively, the result can be interpreted as validating the picture of the slices as thin-wall shells between nearby Euclidean energies carrying an infinitesimal wall tension in a sense made precise by equation (\ref{dthin}). This also lends support to the definition 
\be
\sigma = \int_{\phi_+}^{\phi_0}\sqrt{2(V-V_t)}d\phi \ , 
\ee
for general walls. In terms of the field slope, this corresponds to $\sigma =\int |\dot\phi|d\phi$, which has been advocated elsewhere (e.g. in \cite{Bachas}).

\section*{Acknowledgments\label{sec:ack}} 

I thank  CERN for partial support; Irene Valenzuela and Costas Bachas for discussions that led me to the main idea in this paper 
(ref. \cite{Bachas}, where thick holographic branes are also sliced in thin branes was particularly illuminating); and Pepe Barb\'on, Jeff Fortin, Jes\'us Huertas, Ryusuke Jinno, Thomas Konstandin and Irene Valenzuela for a critical reading of the paper.
This work has been funded by the following grants: IFT Centro de Excelencia Severo Ochoa SEV-2016-0597, CEX2020-001007-S and by PID2019-110058GB-C22 funded by MCIN/AEI/10.13039/501100011033 and by ``ERDF A way of making Europe''.

\appendix

\section{General ${\bma d}$}

\subsection{No Gravity}
The derivation goes through for $d>2$ spacetime dimensions. Without gravity, the starting point is the thin-wall action (see e.g. \cite{Amariti,EF})
\be
S_{\rm thin}=\frac{(d-1)^{d-1}\pi^{d/2}\sigma^d}{\Gamma(1+d/2)\epsilon^{d-1}}\ ,
\ee
where $\sigma$ is the wall tension and $\epsilon>0$ the difference in energy between the two vacua, as in (\ref{thinparam}).
The total thick-wall action is then, using (\ref{dthin}):
\be
S=\int_{\phi_+}^{\phi_0}dS_{\rm thin}=\int_{\phi_+}^{\phi_0}\frac{(d-1)^{d-1}\pi^{d/2}(d\sigma)^d}{\Gamma(1+d/2)(d\epsilon)^{d-1}}=\frac{(d-1)^{d-1}}{\Gamma(1+d/2)}\int_{\phi_+}
^{\phi_0}\frac{[2\pi (V-V_t)]^{d/2}}{(-V_t')^{d-1}}d\phi\ ,
\ee
which reproduces exactly the tunneling action in the $V_t$ formalism. 

\subsection{Gravity On: Minkowski/AdS Vacua}
Including gravity, the thin-wall action for AdS or Minkowski decay
reads
\be
S_{\rm thin}=-\frac{\pi^{d/2}(d-1)^{d-2}}{\kappa \Gamma(1+d/2)C^{d-2}}\left.\left[d\sqrt{1-z}+z(d-1)\,{}_2F_1(1/2,d/2;1+d/2;z)
\right]\right|_{\phi_+}^{\phi_0}\ ,
\ee
where $z=2\kappa_d V_t/C^2$, $\kappa_d=\kappa(d-1)/(d-2)$,
and
\be
 C^2=2\kappa_d V_{t+}+\frac{1}{\sigma^2}\left(\epsilon-\frac12\kappa_d\sigma^2\right)^2=2\kappa_d V_{t-}+\frac{1}{\sigma^2}\left(\epsilon+\frac12\kappa_d\sigma^2\right)^2\ .
 \label{C2d}
\ee
For the infinitesimal slices we have
\be
C^2(\phi)=2\kappa_d V_t+\frac{\epsilon^2}{\sigma^2}=\frac{D_d^2}{2(V-V_t)}\ ,
\label{Cthin}
\ee
with $D_d^2=V_t'{}^2+4\kappa_d(V-V_t)V_t$, 
and the infinitesimal action is
\be
dS_{\rm thin}=
\frac{(d-1)^{d-1}\pi^{d/2}}{C^d\Gamma(1+d/2)}\left[-
\frac{d}{\sqrt{1-z}}+(d-1){}_2F_1(1/2,d/2;1+d/2;z)\right]V_t' d\phi\ .
\ee
Using  the  identity
\be
{}_2F_1(1/2,d/2;1+d/2;z)=\frac{1}{d-1}\left[\frac{d}{\sqrt{1-z}}-(1-z)^{-d/2}{}_2F_1\left(\frac{d-1}{2},\frac{d}{2};\frac{d}{2}+1;\frac{z}{z-1}\right)\right]\ ,
\label{ident}
\ee
we arrive at the action
\be
S=\frac{(d-1)^{d-1}}{\Gamma(1+d/2)}\int_{\phi_+}^{\phi_0}\frac{[2\pi(V-V_t)]^{d/2}}{|V_t'|^{d-1}}{}_2F_1\left(\frac{d-1}{2},\frac{d}{2};\frac{d+2}{2};1-\frac{D_d^2}{V_t'{}^2}\right)d\phi\ ,
\ee
which reproduces the result obtained in the $V_t$ formalism \cite{EF}.

\subsection{Gravity On: dS vacua}
For the dS case, the thin-wall approximation gives the action (for the CdL part) \cite{EF}
\be
S_{\rm thin} = \frac{\pi^{d/2}(d-1)^{d-2}}{\kappa C^{d-2}\Gamma(1+d/2)}\left[2F(z_T)-F(z_+)-F(z_0)+G(z_+)-G(z_T)\right]\ ,
\ee
where
\bea
F(z)&=&d\sqrt{1-z}+z(d-1){}_2F_1(1/2,d/2;d/2+1,z)\ ,\nonumber\\
G(z)&=&2(d-1)z^{1-d/2}{}_2F_1(1/2,d/2;d/2+1,1)\ ,
\eea
with $z=2\kappa_d V_t/C^2$ and $C^2$ is as given in (\ref{C2d}).
The qualitative shape of $V_t$ can be of three different types \cite{EF}, controlled by the value of $z_T$. When the maximum of $V_t$ happens at $\phi_+$ ($\phi_0$), then $z_T=z_+$ ($z_T=z_0$). If the maximum occurs at some point in between, then $z_T=1$. 

For the infinitesimal slices we just need to consider
the two former cases, $z_T=z_+$ and $z_T=z_-$ (corresponding to $V_t'<0$ and $V_t'>0$ respectively). The case $z_T=1$ would occur only for the infinitesimal slice right at the maximum of $V_t$ and is therefore irrelevant for the final result. Using $C(\phi)$ as given in (\ref{Cthin}) and replacing $z_+=2\kappa_dV_t(\phi)/C(\phi)^2$, $z_0=2\kappa_dV_t(\phi+d\phi)/C(\phi)^2$
we get, for the case $z_T=z_+$
\be
dS_{\rm thin}=\frac{\pi^{d/2}(d-1)^{d-2}}{\kappa C^{d-2}\Gamma(1+d/2)}\left[-F'(z)\right]\frac{2\kappa_dV_t'}{C^2}d\phi\ ,
\ee
and for the case $z_T=z_0$
\be
dS_{\rm thin}=\frac{\pi^{d/2}(d-1)^{d-2}}{\kappa C^{d-2}\Gamma(1+d/2)}\left[F'(z)-G'(z)\right]\frac{2\kappa_dV_t'}{C^2}d\phi\ ,
\ee
where $z=2\kappa_dV_t(\phi)/C(\phi)^2$. In both cases, using the identity (\ref{ident}) and paying attention to the sign of $V_t'$, one arrives at
\be
dS_{\rm thin}= \frac{\pi^{(d+1)/2}R_t^d}{\Gamma[(d+1)/2]}(V_t'+|V_t'|)d\phi+
{}_2F_1\left(\frac{d-1}{2},\frac{d}{2};\frac{d}{2}+1;1-\frac{D^2}{V_t'{}^2}\right)s_0 d\phi\ ,
\ee
where
\be
R_t^2=\frac{(d-1)(d-2)}{\kappa |V_t|}\ ,\quad\quad
s_0=\frac{(d-1)^{d-1}[2\pi(V-V_t)]^{d/2}}{\Gamma(1+d/2)|V_t'|^{d-1}}\ .
\ee
This result agrees with the action density obtained in \cite{EF}.

For the Hawking-Moss part of the dS action one has \cite{EF}
\be
S_{\rm HM}=\frac{-4\sqrt{\pi}V}{(d-2)\Gamma\left(\frac{d+1}{2}\right)}\left.\left[\frac{\pi(d-1)(d-2)}{2\kappa V}\right]^{d/2}\right|_{V_+}^{V_{\rm top}}\ .
\ee
The infinitesimal action for a HM transition from $\phi$ to $\phi+d\phi$, obtained taking the $\phi$ derivative of the action above, 
should then be integrated in the HM region of the transition, between $\phi_+$ and some $\phi_{0+}$. One therefore gets 
\be
\delta_{\rm HM}S=\frac{-4\sqrt{\pi}V_t}{(d-2)\Gamma\left(\frac{d+1}{2}\right)}\left.\left[\frac{\pi(d-1)(d-2)}{2\kappa V_t}\right]^{d/2}\right|_{V_{t+}}^{V_{t,0+}}\ .
\ee
This reproduces the corresponding result in the $V_t$ formalism \cite{EF}.

\end{document}